\documentclass[aps,pra,twocolumn]{revtex4-1}
\usepackage{amsfonts}
\usepackage{amssymb}
\usepackage{bm}
\usepackage{array}
\usepackage{color}
\usepackage{graphicx}
\usepackage[pdftex,colorlinks=true,plainpages=false]{hyperref}
\usepackage[utf8]{inputenc}

\begin{document}
\title{Effective Transparency in the XUV: A Pump-Probe Test of Atomistic Laser-Cluster Models}
\author{Rishi Pandit$^1$, Kasey Barrington$^1$, Thomas Teague$^1$,\\  Zachary Hartwick$^1$, Nicolas Bigaouette$^2$, Lora Ramunno$^2$, Edward Ackad$^1$  }
\affiliation{$^1$Department of Physics, Southern Illinois University Edwardsville, Edwardsville, Illinois 62026, USA \\
$^2$Department of Physics, University of Ottawa, Ottawa, Ontario K1N 6N5, Canada}
\date{\today}

\begin{abstract}
 The effective transparency of rare-gas clusters, post-interaction with an extreme ultraviolet (XUV) pump pulse, is predicted by using an atomistic hybrid quantum-classical molecular dynamics model. We find there is an intensity range for which an XUV probe pulse has no lasting effect on the average charge state of a cluster after being saturated by an XUV pump pulse: the cluster is effectively transparent to the probe pulse. The intensity range for which this phenomena occurs increases with cluster size, and thus is amenable to experimental verification. We present predictions for clusters at the peak of the laser pulse profile, as well as the expected experimental time-of-flight signal integrated over the laser profile. Since our model uses only atomic photoionization rates, significant experimental deviations from our predictions would provide evidence for modified ionization potentials due to plasma effects.

\end{abstract}
\maketitle

The extreme ultraviolet (XUV) regime has the simplest interaction between ultra-intense laser pulses and matter, primarily through photoionization. When a nanoscopic dense clump of matter (cluster) is irradiated, secondary ionization events then take place such as collisional ionization. Clusters have solid density but their inter-cluster distance is so large that clusters do not interact with each other, thus they bridge the gap between the gas and solid phases of matter.

Experimental and theoretical laser-cluster interaction studies in the XUV regime are simpler to interpret than at other wavelengths, and it is thus an ideal regime to further test detailed, atomistic models of laser-cluster interactions \cite{Arbeiter2011,Ziaja2013}. At longer wavelengths, even low intensity pulses have efficient processes to transfer energy from the pulse to the free electrons, heating the electron plasma (termed inverse 
Bremsstrahlung heating, IBH) along the axis of the laser's polarization \cite{Peltz2014,Schutte2016}. At shorter wavelengths, the photoionization that occurs is from the inner shell electrons and leads to
subsequent Auger ionization \cite{Hoener2008,Thomas2009,Fennel2010,PhysRevLett.112.183401,Tachibana2015}. 
Thus, XUV pulses -- which through photoionization only access valence shell electrons, and where IBH is negligible for intensities $<10^{16}$~W/cm$^2$ -- present the ideal regime for experiments to probe the degree to which the ionization potential may be modified by the plasma environment
\cite{Gets2006,PhysRevLett.93.043402,PhysRevA.76.043203,Ziaja2013HED,PhysRevA.82.013201,Iwayama2015}.

In this letter, we report on the finding that the ionization in XUV-cluster interaction can become effectively \textit{saturated}. In our model, which uses only atomic ionization potentials, this occurs when a cluster is irradiated with an XUV pulse above a saturation intensity. Additional pulses irradiating the cluster leave no \textit{net} effect on the ionization or total energy; thus the cluster is effectively transparent to the probe pulse.

Multiple models of laser-cluster interaction exist, and new experiments are needed to allow the community to distinguish between the different models \cite{Ziaja2013HED}. Atomistic cluster models to date fall into two primary categories: those with collisional processes beyond single step collisional ionization from the valence shell (atomistic augmented collisional model, AACM)  and those with enhanced photoionization processes arising from ionization potentials that are lowered below the atomic ionization potentials due to the presence of the nanoplasma environment (atomistic augmented photoionization model, AAPM). Our prediction of effective transparency for pump-probe XUV laser cluster interaction was determined using an AACM, and uses only well-established atomic phenomena. Thus an experimental verification of effective transparency would place an upper bound on the significance of enhanced photoionization mechanisms. On the other hand, the failure of the AACM to correctly predict experimental outcomes would be evidence in favor of an enhanced photoionization mechanism (such as, eg, electron screening or barrier suppression). Thus this letter presents a proposal for an experiment.

The schema to distinguish the two models is as follows. A pump pulse irradiates a cluster at an intensity that is above the saturation intensity predicted by the AACM, where atomic ionization potentials are used. Above the saturation intensity, the average ion charge state (AICS, the total charge divided by the number of ions since only ions are detectable) of the cluster in the calculation is independent of pump intensity. This occurs when the intensity is high enough that all possible photoionizations have occurred, but not high enough for significant multiphoton photoionization or IBH. The AACM predicts that the cluster would then be effectively saturated, and would not increase its AICS if subsequently irradiated by a probe pulse.

However, if the electromagnetic fields of the nanoplasma sufficiently perturb the atomic ionization potentials so that single-photon ionization from deeper states becomes possible, effective transparency would not be detected in an experiment. The strength of the nanoplasma perturbation would have to be a function of the plasma density. If a probe pulse irradiates the cluster after only a short delay, while the cluster is still dense, the different models will strongly disagree. The nanoplasma perturbation, if large, would allow the cluster to further ionize due to the lowered ionization potentials. With increasing delay of the probe pulse the cluster's density decreases and so does the nanoplasma's perturbation. Thus, the two models predict different trends as a function of pump-probe delay.

This methodology is complimentary to previous proposals \cite{Ziaja2013} with the advantage that the effect is enhanced by the cluster size distribution. All previous work in this area has neglected the role of collisional excitation which is known to play a dominant role in the ionization in the XUV \cite{Muller2015}.

Our implementation of an AACM is a hybrid approach wherein the particles are treated as classical charge distributions
whose motion is solved by molecular dynamics. The ionization rates are determined from a mix of experimental (when
available) and theoretical cross-sections in the gas phase \cite{AckadPRL}. 
During ionization, the perturbation on the ions due to the cluster environment (the nanoplasma) has been shown to be well represented by our Local Ionization Threshold (LIT)
model \cite{ionization_approx}, which maintains the use of atomic ionization potentials. Using the LIT model we include single- and multi-photon ionization, collisional ionization, augmented collisional ionization (ACI) \cite{AckadPRL}, and
many-body recombination \cite{RecombAckad}.
The model has been successful in reproducing the laser-cluster experimental signals 
\cite{Ackad_cluster_expansion_XUV}, including experiments where Auger ionization is dominant \cite{RecombAckad}.

In AACMs, collisional ionization beyond a single step process is considered. The standard ionization channels are augmented to include the possibility of collisional 
excitation, so called augmented collisional ionization (ACI) \cite{AckadPRL}. 
A bound electron can first be promoted from the ground state to an excited state by a 
collision of an already ionized electron. Subsequently, this excited electron can be
ionized by being promoted from the excited state to the continuum through a second collision. While the \textit{whole trip} can be 
energetically the same, breaking the process up into two steps reduces the energy required for each transition. 
This allows an electron with less kinetic energy (compared with single-step ionization) to 
execute the process. In a nanoplasma, the energy distribution is, on average, Maxwellian and 
thus there are many more electrons with enough energy to excite an atom than there are who 
can ionize an atom directly \cite{Ackad_cluster_expansion_XUV}. This ionization pathway leads to higher charge states in 
the cluster and collisionally reduced photoabsorption (CRP) where clusters absorb less photons due to fast collisional ionization removing
target ions \cite{Ackad_cluster_expansion_XUV}.

The current work includes one and two photon ionization given by the rate,
\begin{equation}
 \frac{dN}{dt} = \left(\frac{I}{E_{ph}}\right) \sigma^{(1)} + \left(\frac{I}{E_{ph}}\right)^2 \sigma^{(2)}
\end{equation}
where $I$ is the intensity of the laser, $E_{ph}$ is the photon energy and $\sigma^{(n)}$ is the 
$n$-th order photoionization process. 
The values of $\sigma^{(1)}=5.0\times 10^{-18}$ cm$^2$
 \cite{Argon_photo_neutral} and 
$\sigma^{(2)}= 10^{-50}$ cm$^4$/s (taken as an upper limit from reference~\cite{McKenna2004}) were used. 
The higher $\sigma^{(2)}$ is, the smaller the range of intensities in which effective saturation will occur, and thus taking an upper limit
gives a conservative estimate of the saturation effect.

To show the saturation effect in our AACM model, we solved the interaction of argon  clusters (Ar$_{147}$) irradiated by two XUV pulses at $\lambda =33$ nm (37.6 eV) 25-fs apart. 
Both pulses have a full-width-at-half-maximum of 10 fs. 
Although short pulses increase the probability of multiphoton ionization (which undermines our signal), they also allow the probe pulse to 
irradiate the cluster while the density is still high (which enhances the likelihood of nanoplasma
perturbations which must depend on the plasma density). The number density of the ions at the peak of the pump pulse is around $3.99\times10^{-3}$ bohr$^{-3}$ 
(where the distance of the furthest ion is used as the radius of the spherical volume) while at the peak
of the probe pulse the density is $3.84\times10^{-3}$ bohr$^{-3}$ , a percent difference of about 3.85\%. Further,
the plasma number-density of the cluster at the same radius is $1.27\times10^{-3}$ bohr$^{-3}$  at the peak of the pump and $1.58\times10^{-3}$ bohr$^{-3}$  at the peak of the probe. 
Thus, the effects of an enhanced photoionization mechanism will be most pronounced during the probe pulse and would decrease as the pulse delay increases.



The specific XUV-wavelength was chosen to be above 
the singly ionized ionization potential for argon 
(27.6 eV) and below any significant inner-ionization thresholds. An intensity scan was then 
performed for the pump pulse. The AICS after 500~fs of the start of the pump pulse is used as a measure of the overall ionization of the cluster, and the cluster is considered to be at the focus of the laser pulse.
The average is taken over all ions; ions containing classically bound electrons 
have their charges decreased accordingly. 

\begin{figure}[htbp]
\begin{center}
\includegraphics[width= 8.0cm]{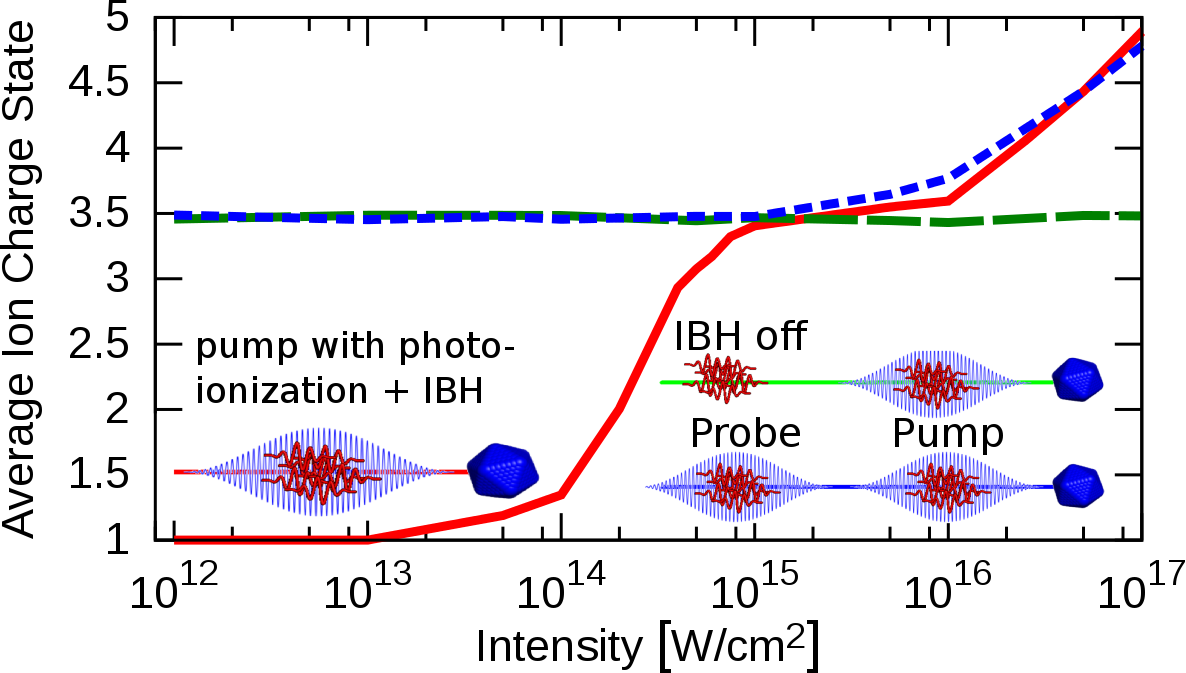}
\caption{(Color online) The solid (red) curve shows average ion charge state (AICS) vs pump intensity for Ar$_{147}$ irradiated by a single 10 fs, $\lambda=33$~nm pump pulse (depicted in the lower left illustration).  
The short-dashed (blue) curve shows the AICS vs probe intensity for Ar$_{147}$ irradiated by a pump pulse fixed at an intensity of $2.5\times 10^{15}$~W/cm$^2$ and subsequently (25 fs delay) 
irradiated by probe pulse of varying intensity (depicted by the lower right illustration). The long-dashed (green) curve shows the AICS vs probe intensity for the same setup, except where the probe pulse can only photoionize and thus no IBH occurs (depicted by the top right illustration). }
\label{fig_intensity_vs_charge_states}
\end{center}
\end{figure}

The solid red curve in Fig.~\ref{fig_intensity_vs_charge_states} shows the AICS vs pump intensity when an Ar$_{147}$  cluster is irradiated only by a pump pulse. As pump intensity is increased from $10^{12}$~W/cm$^2$ to $10^{17}$~W/cm$^2$, the AICS starts to increase very gradually. At around $10^{14}$ W/cm$^2$ the AICS increases dramatically until, at an intensity of about $10^{15}$~W/cm$^2$, the AICS becomes saturated around AICS=3.5. This is what we call the "saturation intensity". Further increasing the pump intensity, only marginally increases the AICS until after the AICS plateau, around $10^{16}$~W/cm$^2$. 
This small increase is due to multiphoton ionization and IBH. At an intensity of 
$10^{17}$~W/cm$^2$ AICS reaches about 4.9. Even at this intensity, more than 50\% of the ionization is due to collisional ionization, almost exclusively through ACI. 
At the saturation intensity ACI accounts for well above 90\% of all ionizations. 

To demonstrate effective saturation, a pump probe setup is modeled showing that the probe pulse has almost no effect on the AICS.
The pump pulse is fixed at $2.5\times10^{15}$~W/cm$^2$, just above the saturation intensity.
The intensity of the subsequent probe pulse (25~fs later) is 
scanned from $10^{12}$~W/cm$^2$ to $10^{17}$~W/cm$^2$ as depicted at the bottom right illustration of Fig.~\ref{fig_intensity_vs_charge_states}.
The blue short-dashed curve in Fig.~\ref{fig_intensity_vs_charge_states}) shows AICS versus probe intensity, where the AICS is measured 500~fs after the 
start of the pump pulse. We find that the AICS begins and remains saturated until 
the intensity of the probe pulse exceeds about $10^{16}$~W/cm$^2$. This is when the probe pulse reaches sufficient intensity for IBH to become significant.
Below this intensity, the additional probe pulse does not meaningfully increase the AICS from what it was after the pump; this is the basis for terming the phenomenon
effective transparency, since it is as if the cluster were transparent to the probe pulse.


Why is there a plateau in the AICS? An analysis of the charge state distribution verses time shows that the irradiation of the cluster by the pump pulse at the
saturation intensity ionizes all possible targets via photoionization and collisional ionization. Thus, during the pulse there are no more targets to
further photoionize \cite{Rishi_coming}. This is the high intensity limit to the previously observed CRP \cite{Ackad_cluster_expansion_XUV}. Without any targets the probe pulse
does not contribute to the AICS. The result is thus the same saturated AICS, both with and without the probe pulse. 
Conceptually, this is where any enhanced photoionization mechanisms would play a significant role. The probe pulse 
is irradiating a dense nanoplasma and, according to ionization potential (IP) lowering models,
would allow for the photoionization of ions well beyond Ar$^{1+}$ and
thus change the final AICS significantly \cite{PhysRevLett.93.043402,PhysRevA.82.013201}. 

Our simulations further show that the effective transparency phenomenon is fairly insensitive to the change of the delay time from 15~fs to 
around 150~fs. Noticeable deviations occur only when the delay time is $>$200~fs. 
This insensitivity would be a verifiable trend in the experimental data only if no significant ionization potential lowering occurs. 

As the density of the cluster decreases with the cluster's disintegration, one would expect the IP lowering effect must also decrease. It would tend to zero as the density becomes that of a gas, since no IP lowering mechanism has been observed in gas \cite{PhysRevLett.93.043402,georgescu:043203,murphy:203401,PhysRevA.83.043203,Ziaja2013HED}. 
IP lowering effects would thus be sensitive to pump-probe delay time.
If a lack of sensitivity to the delay time (within the 15-150~fs range for the aforementioned parameters) were found experimentally, it would place constraints on how strong IP lowering contributes to the total ionization.

Artificially turning off the probe pulse's electric field, allowing only direct photoionization the cluster (no IBH), shows that the end of the AICS plateau is due almost exclusively to IBH (short-dashed green curve in figure~\ref{fig_intensity_vs_charge_states}).

\begin{figure}[htp]
\begin{center}
\includegraphics[scale=0.20]{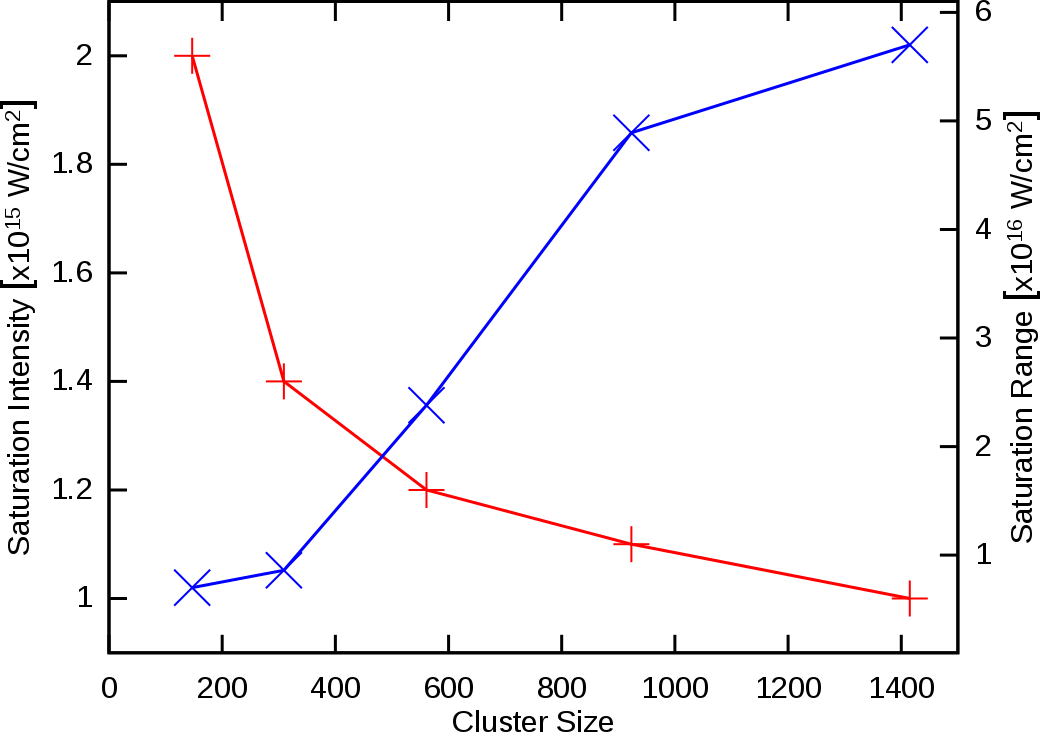}
\caption{(Color Online) The saturation intensity (minimum intensity needed to saturate the cluster) $I_{sat}$ as a function of the cluster size is shown as the (red) pluses for pump pulse duration of 10~fs at $\lambda=33$ nm. The intensity range (right vertical axis) as a function of cluster size over which the probe pulse has a negligible effect on the average ion charge state is shown as the (blue) x's.}
\label{sat_intensity}
\end{center}
\end{figure}

We now consider calculations that correspond more directly to what an experiment would detect. In any cluster beam, there is a log-normal distribution of cluster sizes. Thus, we examine the effect of cluster size on saturation intensity, and the intensity range over which the AICS remains constant.
In the range of parameters examined, the saturation intensity $I_{sat}$ decreases as the cluster size increases (shown as the red plus signs in Fig.~\ref{sat_intensity} where the line is drawn to aid the eye). 
This makes intuitive sense since the larger clusters absorb the same amount of energy \textit{per ion} as the smaller clusters. However, the amount of energy needed for an electron to escape the cluster remains the same \cite{Arbeiter2011}. 
Thus, larger clusters absorb more \textit{total} energy (than smaller clusters) at the same intensity. It thus takes less intensity to effectively saturate the cluster's ionization channels. It should be noted that the trend ends once the cluster's size becomes large enough that the pulse is significantly depleted by the photoabsorption ($N \ge 2057$).

The range of intensities over which the cluster is effectively transparent ($I_{\rm high} - I_{\rm low}$) to the probe pulse also increases with the size of the cluster (shown as the blue x's in Fig.~\ref{sat_intensity}). 
This indicates that the effective saturation is more pronounced in all measures in larger clusters. It was further found that AICS $\approx \alpha \ln(\beta N)$, where $\alpha =0.227$, $\beta=35457.1$ and $N$ is the cluster size less than 2057 for argon \cite{Rishi_coming}. The cluster size distribution will change the AICS but not by much due to the logarithmic relationship between AICS and $N$. Thus, experiments with the cluster size peaked at a few hundred atoms would have their signals enhanced by the cluster beam's size distribution. 



\begin{figure}[tbp]
\begin{center}
\includegraphics[width= 8.0cm]{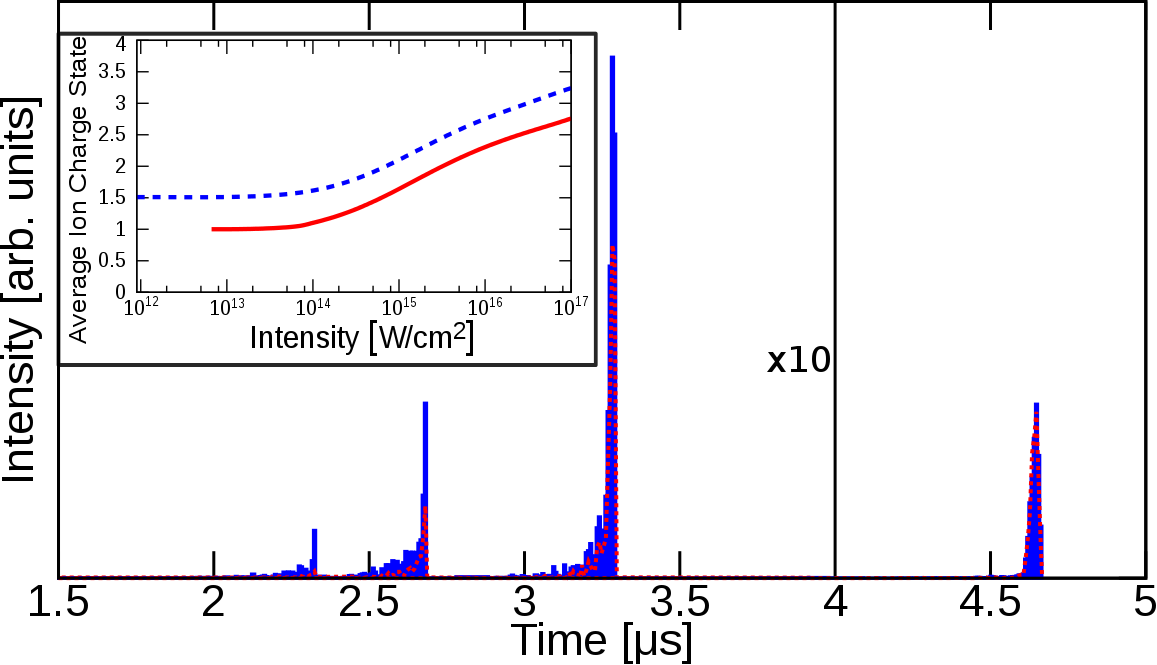}
\caption{(Color Online) Time-of-flight signal for Ar$_{147}$ irradiated only by a 10~fs pump pulse of $I=2.5\times10^{15}$~W/cm$^2$ at $\lambda=33$~nm shown as the dashed red line. The blue boxes show the time-of-flight signal when the pump pulse is followed by an identical probe pulse. 
Inset: The average ion charge state as a function of the pump pulse's intensity (solid red) integrated over the full laser pulse's spatial distribution. The dashed (blue) curve is the average ion charge state for a pump pulse fixed at $2.5\times 10^{15}$~W/cm$^2$ as a function of the 
probe pulse's intensity, integrated over the spatial profiles of the two pulses. }
\label{fig_tof1}
\end{center}
\end{figure}
Thus far the results have been for the spatial peak of the laser pulse(s) and may be achievable if the beam is masked to reduce the wings of the pulse as in Ref.~\cite{Hoener2008}. Otherwise, we now consider what an experiment would detect due to the spatial distribution of the pulse. While a small subset of clusters will be irradiated by both pulses
at the peak intensities, many clusters spatially located in the wings of the pulse will be irradiated by a pump pulse of insufficient intensity to saturate the ionization channel. 
The probe pulse will then increase their ionization. 
In the pump-probe 
setup, clusters were assumed to interact with the same intensity region of both pulses, i.e., the pulses were assumed to be spatially identical and focused at the same location.
The resulting time-of-flight (TOF) signal for the pump pulse alone at $I=2.5 \times10^{15}$~W/cm$^2$ is shown as the dashed red line in Fig.~\ref{fig_tof1}. 
It was calculated using the methodology from reference~\cite{RecombAckad}, where the signal is integrated over the intensities of the pulse for a single cluster size and using the TOF setup described in reference~\cite{Thomas2009}). 
It shows that the signal will contain primarily singly charged ions with an almost linear decrease in the higher charge states. The signal only sees a small change when an identical probe pulse is included (shown as the blue boxes in Fig.~\ref{fig_tof1}). If the probe pulse is below the saturation intensity (but with the same spatio-temporal profile), the TOF is quite close to the pump-only signal. However, increasing the probe pulse to beyond the saturation intensity 
results in some increase in the signal from the multiply-charged states. This is to be expected as now more clusters will fall into a spatial region where they will be saturated by the probe pulse.

To illuminate this effect and show the trends an experiment would see in the absence of IP lowering, the AICS is calculated over the entire spatial distribution of the pulse. The saturation of the AICS is not observable for a single pulse (red solid curve in the inset of Fig.~\ref{fig_tof1}) due to the spatial wings of the pulse. However, saturation {\it is} observable when the clusters are further irradiated by a probe pulse. Fixing the pump pulse's peak intensity at $2.5\times10^{15}$~W/cm$^2$ and changing the intensity of the probe pulse shows the saturation of the AICS. As the probe pulse's intensity increases from $10^{12}$ to about $10^{14}$~W/cm$^2$, the AICS remains constant at about 1.5 (blue dashed curve in the inset of Fig.~\ref{fig_tof1}). Further increases in the intensity of the probe pulse increase the AICS as more clusters in the wings of the pulse become saturated. This result is again constant with delay times less than about 200~fs. IP lowering models would show much sharper increases in the AICS as a function of the intensity. This would result in the AICS increasing even for a low intensity probe pulse since the additionally photoionized electrons, allowed by IP lowering, would be cluster bound causing additional collisional ionization events.
%

In conclusion, we have shown that atomic-based laser-cluster interaction models predict that it is possible to induce effective transparency in the XUV using a pump-pulse setup. This effect is insensitive to the delay between the pulses, and thus insensitive to nanoplasma density; this is in contrast to what IP lowering models would predict. Experimental verification of our results would place strict limits on the role of IP lowering mechanisms for small rare-gas clusters and would provide the field with valuable data to refine its models of photoionization in laser-cluster interactions not only in the XUV, but for any wavelength where photoionization plays a major role.

E. A. would like to thank M. M\"{u}ller for useful discussions. This work was supported by Air Force Office of Scientific Research under AFOSR Award No. FA9550-14-1-0247.
\bibliographystyle{unsrt}
\bibliography{photobleaching}

\end{document}